\documentclass[%
 reprint,
 amsmath,amssymb,
 aps,
prl,
]{revtex4-2}

\usepackage{graphicx}
\usepackage{dcolumn}
\usepackage{bm}
\usepackage{hyperref}
\usepackage{physics} 
\usepackage{siunitx} 
\usepackage{xcolor} 
\usepackage{xspace}

\newcommand*{\ind}[1]{\ensuremath{\kern.001em_\mathrm{#1}}}  
\newcommand{\ie}{i.\,e.\xspace}

\begin{document}

\title{Minority-Triggered Reorientations Yield Macroscopic Cascades and Enhanced Responsiveness in Swarms}

\author{Simon Syga}
 \email{simon.syga@tu-dresden.de} 
 \affiliation{Information Services and High Performance Computing, Center for Interdisciplinary Digital Sciences, TUD Dresden University of Technology, 01062 Dresden, Germany}
\author{Chandraniva Guha Ray}
 \affiliation{Max Planck Institute for the Physics of Complex Systems, N\"othnitzer Stra\ss e 38, 01187 Dresden, Germany}
\affiliation{\smash{Max Planck Institute of Molecular Cell Biology and Genetics, Pfotenhauerstra\ss e 108, 01307 Dresden, Germany}}
\affiliation{Center for Systems Biology Dresden, Pfotenhauerstra\ss e 108, 01307 Dresden, Germany}
 \affiliation{Department of Physical Sciences, Indian Institute of Science Education and Research Kolkata, Mohanpur, 741246, India}
\author{Josué Manik Nava Sedeño}
 \affiliation{Departmento de Matemáticas, Facultad de Ciencias, Universidad Nacional Autónoma de México, Circuito Exterior, Ciudad Universitaria, 04510 Mexico City, Mexico}
\author{Fernando Peruani}
 \affiliation{Laboratoire de Physique Théorique et Modélisation, UMR 8089, CY Cergy Paris Université, 95302 Cergy-Pontoise, France}

\author{Andreas Deutsch}
 \affiliation{Information Services and High Performance Computing, Center for Interdisciplinary Digital Sciences, TUD Dresden University of Technology, 01062 Dresden, Germany}

\date{\today}

\begin{abstract}
Collective motion in animals and cells often exhibits rapid reorientations and scale-free velocity correlations.
This allows information to spread rapidly through the group, allowing an adequate collective response to environmental changes and threats such as predators.
To explain this phenomenon, we introduce a simple, biologically plausible mechanism:
a minority-triggered reorientation rule. 
When local order is high, agents sometimes follow a strongly deviating neighbor instead of the majority.
This rule qualitatively changes the macroscopic system behavior compared to traditional flocking models, as it generates heavy-tailed cascades of reorientations over broad parameter ranges.
Our mechanism preserves cohesion while markedly enhancing collective responsiveness because localized directional cues elicit amplified group-level reorientation.
Our results provide a parsimonious, biologically interpretable route to critical-like fluctuations and high responsiveness during flocking.
\end{abstract}

\maketitle
\textit{Introduction}---Collective motion is ubiquitous across biological scales, from cellular migration in embryonic development and cancer invasion \cite{chepizhko2016, peruani2012, wood2023, deutsch2020} to aerial displays of starling flocks and evasive maneuvers of fish schools \cite{cavagna2010, procaccini2011, handegard2012}. 
The synchronization of movement in these diverse systems exhibits remarkably similar macroscopic patterns, suggesting universal organizing principles that transcend specific biological details \cite{castellano2009, vicsek2012}.

Empirical studies of collective animal motion have documented scale-free velocity correlations in starling flocks \cite{cavagna2010, bialek2012a} and large density fluctuations that result in scale-free avalanches of regrouping events in sheep flocks \cite{ginelli2015}.
This results from velocity fluctuations that propagate over distances many times larger than the range of direct individual interactions \cite{cavagna2010}. 
In starling flocks, for example, correlation lengths scale linearly with system size, so velocity fluctuations propagate over distances many times greater than the range of direct individual interactions \cite{cavagna2010}. 
This allows information, such as the presence of predators, to travel rapidly throughout the group through avalanches of directional change \cite{procaccini2011, rosenthal2015, gomez-nava2023}. 
Such critical-like behavior was hypothesized to confer adaptive advantages by optimizing the trade-off between group cohesion and rapid response to threats \cite{mora2011, cavagna2018, klamser2021, ouellette2022}.

Mathematical models of collective motion, predominantly based on the seminal Vicsek model (VM) \cite{vicsek1995}, have successfully reproduced many features of biological flocks \cite{couzin2002,vicsek2012}. 
In the VM, self-propelled agents align their motion with that of local neighbors under the influence of noise. 
As noise decreases or density increases, the system undergoes a transition from disordered to ordered collective motion \cite{vicsek1995,chate2008,ginelli2016}. 
Yet, in standard flocking models in the ordered phase, the response to perturbations is weak and localized. 
This raises the question: How do natural collectives achieve strong, critical-like responsiveness to local perturbations \cite{mora2011}?

\begin{figure*}
\centering
\includegraphics{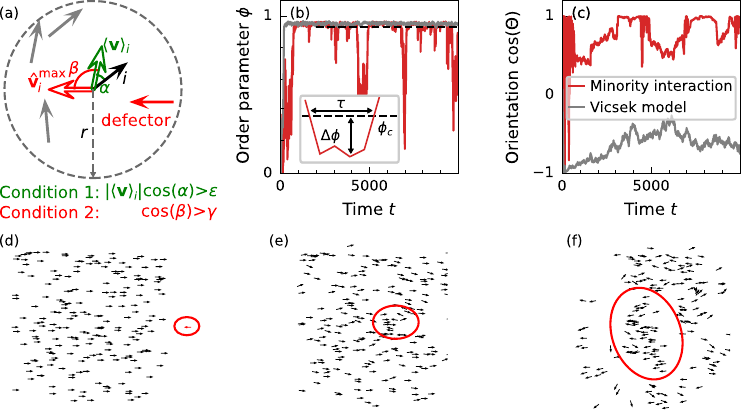}
\caption{\label{fig:minority_interaction} \textbf{Minority interaction leads to avalanches.} (a) In standard Vicsek dynamics, a particle $i$ (black) aligns with the average orientation of its neighbors (green). 
With the minority interaction, if the surrounding neighborhood is sufficiently ordered (condition 1), the same particle may instead follow a defector (red) that strongly deviates from the local consensus (condition 2). 
The outlined red arrow indicates the focal particle's orientation after reorienting due to the minority interaction. 
 (b) The minority interaction creates perturbations that propagate through the system as an avalanche of reorientations, temporarily disrupting global order. 
 Time evolution of the order parameter $\phi(t)$ in our model with minority interaction ($L=32, \rho=1, \gamma=-0.3, \epsilon=0.6, \eta=0.1$), showing large fluctuations corresponding to avalanche events. 
 The dashed line ($\phi_c$) indicates the threshold used to define avalanche duration $\tau$ and size $\Delta \phi$. 
The duration of an avalanche $\tau$ is defined as the time period during which the order parameter is below this threshold and the avalanche size is the maximum deviation of the order parameter from $\phi_c$, see inset. 
In contrast, the order parameter for the classical Vicsek model with identical parameters (gray line) shows much smaller fluctuations. 
(c) The direction of the average velocity, $\cos(\Theta(t))$, also exhibits large fluctuations coinciding with the avalanches in $\phi(t)$ for the minority model (red), whereas it remains relatively stable for the standard VM (gray).
(d)-(f) Snapshots of an example simulation, where a single defector (red circle in d) causes an avalanche of reorientations (red circles in e and f) in an initially ordered flock.}
\end{figure*}

\textit{Model definition}---We address this puzzle by augmenting the VM \cite{vicsek1995} with a biologically motivated minority interaction: 
in well-aligned neighborhoods, a particle may briefly abandon majority consensus to follow the most strongly deviating neighbor (defector), as observed during predator evasion in fish, birds, and sheep \cite{procaccini2011,handegard2012,marras2013,rosenthal2015,gomez-nava2023,ginelli2015}.
The resulting competition between conformity and deviation triggers avalanche-like cascades of reorientations that propagate through the group before standard alignment restores coherence, yielding critical-like heavy-tailed event statistics and long-range correlations without fine tuning.

We consider $N$ point particles moving with constant speed $v_0$ in a two-dimensional square domain of length $L$ with periodic boundary conditions. 
Each particle $i$ is characterized by its position $\vb{x}_i$ and orientation $\theta_i \in [0, 2\pi)$, with the corresponding unit velocity vector $\vu{v}_i = (\cos\theta_i, \sin\theta_i)^T$. 
Particles interact with neighbors within a circular neighborhood with fixed radius $r$.

The dynamics proceeds in discrete time steps $\Delta t$ through two steps: orientation update and position update. 
The orientation update incorporates both the standard majority-based alignment rule from the VM and our novel minority interaction mechanism (Fig.~\ref{fig:minority_interaction}a).

To formalize the minority interaction, we define two quantities for each particle $i$ at time $t$. First, the local flux vector:
\begin{equation}
    \ev{\vb{v}}_i(t) = \frac{1}{N_i} \sum_{j, |\vb{x}_i - \vb{x}_j| < r} \vu{v}_j(t)
    \label{eq:local_flux}
\end{equation}
where $N_i$ is the number of neighbors within a distance $r$ (including particle $i$ itself). 
This vector captures the average movement direction in the local neighborhood. 
Second, we identify the defector---the neighboring particle $k$ whose orientation $\vu{v}_k(t)$ deviates most strongly from the local consensus:
\begin{equation}
    \vu{v}_i^{\text{max}} = \vu{v}_k \quad \text{where} \ k \in \underset{j, |\vb{x}_i - \vb{x}_j| < r}{\text{argmin}} \left( \vu{v}_j \cdot \ev{\vb{v}}_i \right),
    \label{eq:defector}
\end{equation}
for a fixed time $t$.
The minority interaction is triggered when two conditions are simultaneously satisfied:
(i) The focal particle's neighborhood exhibits sufficient local alignment,
\begin{equation}
    \ev{\vb{v}}_i(t) \cdot \vu{v}_i(t) > \epsilon,
    \label{eq:condition_alignment}
\end{equation}
where $\epsilon \in [-1, 1]$ is a threshold parameter. 
Higher values of $\epsilon$ require stronger local consensus before the minority interaction becomes possible. 
This condition reflects that only individuals within a coherent group are sensitive to a starkly deviating member.
Note that without this condition, there would be no clear defector since there is no majority to deviate from.
(ii) The defector's orientation sufficiently contradicts the local consensus,
\begin{equation}
    \ev{\vb{v}}_i(t) \cdot \vu{v}_i^{\text{max}}(t) < \gamma,
    \label{eq:condition_deviation}
\end{equation}
where $\gamma \in [-1, 1]$ is a second threshold parameter. 
Lower values of $\gamma$ require more extreme deviation by the defector.
When both conditions (Eqs.~\ref{eq:condition_alignment} and \ref{eq:condition_deviation}) are satisfied, the focal particle adopts the defector's orientation (with noise):
\begin{equation}
    \theta_i(t + \Delta t) = \theta_k(t) + \xi_i(t)
    \label{eq:minority_update}
\end{equation}
where $k$ is the index of the defector and $\xi_i(t)$ is Gaussian noise with standard deviation $\sigma$.
Otherwise, the particle follows the standard Vicsek alignment rule
\begin{equation}
    \theta_i(t + \Delta t) = \arg[\ev{\vb{v}}_i(t)] + \xi_i(t)
    \label{eq:vicsek_update}
\end{equation}
After orientation updates, positions are updated according to
\begin{equation}
    \vb{x}_i(t + \Delta t) = \vb{x}_i(t) + v_0 \vu{v}_i(t + \Delta t) \Delta t
    \label{eq:position_update}
\end{equation}

\textit{Numerical results}---We systematically characterize collective behavior across a broad parameter space.
Following standard practice \cite{chate2008, ginelli2016}, we set $\Delta t = r = 1$ and fix particle speed at $v_0 = 0.5$. 
We systematically explore:
(i) system size: $L \in \{16, 32, 64, 128\}$ with periodic boundaries,
(ii) particle density: $\rho = N/L^2 \in \{0.5, 1.0, 1.5\}$ corresponding to $N\in[128, 24576]$ across explored system sizes,
(iii) noise amplitude: $\eta \in \{0.1, 0.2, 0.3\}$, with noise standard deviation $\sigma = \eta 2 \pi / \sqrt{12}$ (chosen for comparison with uniform noise used in some VM studies \cite{chate2008}),
(iv) minority interaction thresholds: $\epsilon \in [0, 1]$ and $\gamma \in [-1, 0]$.

For each parameter combination, we perform long simulations ($10^6$ time steps, excluding the initial $10^4$ time steps, more than 100 times the correlation time of the order parameter, to rule out transient behavior \cite{chate2008}).

To isolate the effects of minority interaction, we conduct parallel simulations of the standard VM using identical parameter values.

We quantify the collective behavior using the standard polar order parameter
\begin{equation}
    \phi(t) := \frac{1}{N} \left| \sum_{i=1}^N \vu{v}_i(t) \right|.
    \label{eq:order_parameter}
\end{equation}
This parameter measures global alignment, ranging from $\phi = 1$ for perfect alignment to $\phi \approx 0$ for disordered motion (in large systems). 
The order parameter is our primary observable for identifying and characterizing avalanche dynamics.
We also analyze the dynamics of the average velocity direction
\begin{equation}
    \Theta(t) := \arg \left( \sum_{i=1}^N \vu{v}_i(t) \right).
    \label{eq:flux_direction}
\end{equation}

We observe large-scale fluctuations of the order parameter $\phi$ corresponding to avalanches caused by defector particles (Fig.~\ref{fig:minority_interaction}b). 
In contrast, the VM with corresponding parameters shows much smaller fluctuations that scale approximately as $1/\sqrt{N}$, as expected for finite-size effects in the ordered phase \cite{chate2008}. 
We find that avalanches in global order coincide with large-scale fluctuations of the global velocity direction (Fig.~\ref{fig:minority_interaction}c), indicating that the minority interaction enables macroscopic changes in collective behavior.

Before investigating the system's behavior in equilibrium, we first directly demonstrate the enhanced responsiveness of a flock in our model. 
To this end, we subject an initially ordered flock to a controlled perturbation---a single particle forced to orient opposite the collective for several time steps, then released---and find that in this protocol the minority model produces amplified, group-wide reorientation while the VM rapidly realigns (Fig.~\ref{fig:minority_interaction}d--f, Fig.~\ref{fig:s1_responsiveness}).

Next, we systematically study the quasi-equilibrium behavior of the system, starting from disordered initial conditions.
To analyze the statistical properties of the observed avalanches, we examine two key characteristics: the avalanche size and the avalanche duration. 
For temporal dynamics, we record distributions of ordered and disordered periods of the system. 
We define the system to be in an avalanche state at time $t$ if $\phi(t) \le \phi_c$, where $\phi_c := \ev{\phi}_{\text{VM}} - 3 \, \text{STD}(\phi_{\text{VM}})$ is a threshold defined by the mean order parameter and its standard deviation in the corresponding classical VM simulation. 
This allows us to define the avalanche size of avalanche $j$ as the maximum deviation from this threshold, \ie, $\Delta\phi_j := \max_{t \in I_j} (\phi_c - \phi(t))$, where $I_j = [t_{\text{start},j}, t_{\text{end},j}]$ is the time interval associated with avalanche $j$ (when $\phi(t) \le \phi_c$). 
Moreover, we define the return time $\tau$ as the duration of these disordered periods ($\tau_j := t_{\text{end},j} - t_{\text{start},j}$).
See the inset in Fig.~\ref{fig:minority_interaction}b for a sketch of the avalanche observables.

\begin{figure}
\centering
\includegraphics{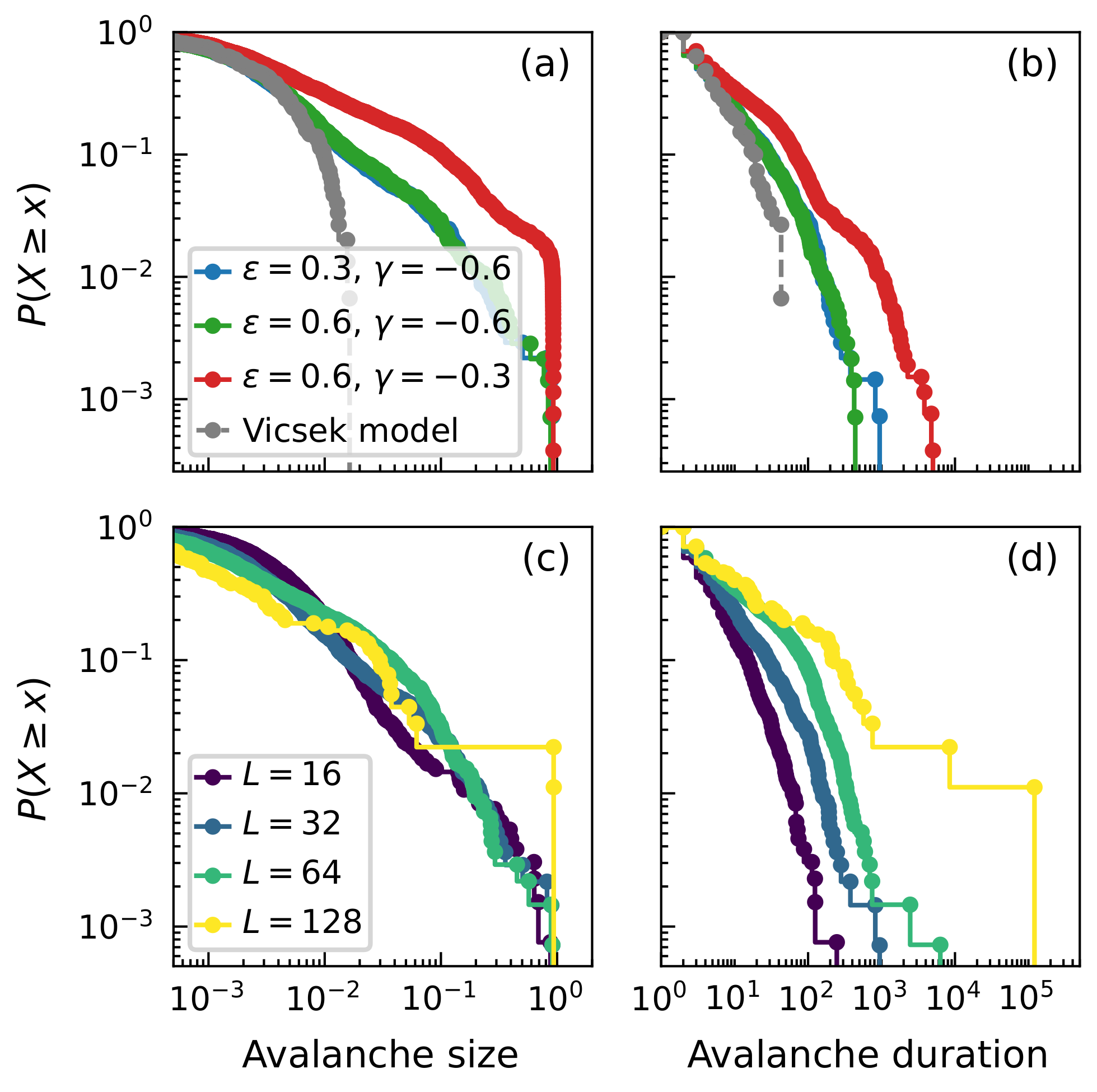} 
\caption{\label{fig:avalanche_stats} \textbf{Avalanche statistics.} Complementary cumulative distribution functions (CCDF), $P(X \ge x)$. (a) Avalanche size $\Delta\phi$ for different minority interaction thresholds ($\epsilon, \gamma$) compared to the standard VM (gray). System parameters: $L=32, \rho=1.0, \eta=0.1$. (b) Avalanche duration (return time $\tau$) for the same parameters as (a). (c) Avalanche size CCDF for fixed thresholds ($\epsilon=0.3, \gamma=-0.6$) and varying system sizes $L$. (d) Avalanche duration CCDF for the same parameters as (c). Avalanches can be macroscopic and orders of magnitude larger than in the standard VM.}
\end{figure}

In large parameter ranges of the threshold parameters $\epsilon, \gamma$, we find macroscopic avalanches that are orders of magnitude larger (Fig.~\ref{fig:avalanche_stats}a) and longer-lasting (Fig.~\ref{fig:avalanche_stats}b) than in the VM. 
The distribution of avalanche durations exhibits heavy tails over at least three orders of magnitude. 
This temporal behavior persists across the explored system sizes (Fig.~\ref{fig:avalanche_stats}c,d and Fig.~\ref{fig:s2_avalanches}).

Next, we investigate the effect of the minority interaction on the collective spatial dynamics, particularly the local transmission of directional information. 
To this end, we calculate the correlation of velocity fluctuations $\Delta\vu{v}_i := \vu{v}_i - \ev{\vu{v}}$ of particle pairs at distance $d$,
\begin{equation}
    C(d) := \frac{\sum_{i, j} \Delta\vu{v}_i \cdot \Delta\vu{v}_j \, \delta(d_{i,j} - d)}{\sum_{i, j} \delta(d_{i,j} - d)},
    \label{eq:correlation_func}
\end{equation}
where $d_{i,j} = |\vb{x}_i - \vb{x}_j|$,  $\ev{\vu{v}} = 1/N \sum_i \vu{v}_i$ is the global average velocity, and $\delta(x)$ is the Dirac delta function. 
We average this observable over 1000 time points chosen uniformly randomly after the relaxation time.

\begin{figure}
\centering
\includegraphics{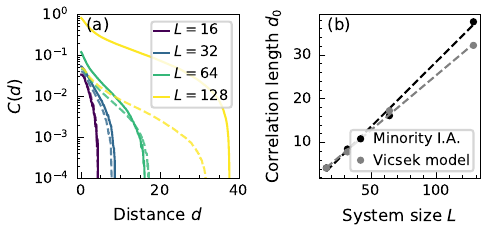}
\caption{\label{fig:correlation} \textbf{Spatial velocity correlations.} (a) Velocity fluctuation correlation function $C(d)$ versus inter-particle distance $d$ for systems with different sizes $L$ (indicated by color). Solid lines: model with minority interaction. Dashed lines: corresponding standard VM. Correlations are significantly stronger with the minority interaction. (b) Correlation length $d_0$ (defined as $C(d_0)=0$) scales linearly with system size $L$ for the minority model (black circles) and the VM (gray circles). Dashed lines are linear fits. Model parameters identical to those in Fig.~\ref{fig:avalanche_stats}c,d.}
\end{figure}

The minority model shows significantly stronger short-range velocity correlations than the VM; for $L=128$, the difference is tenfold (Fig.~\ref{fig:correlation}a and Fig.~\ref{fig:s3_correlationfunction}).
The correlation length $d_0$, defined as the distance where $C(d_0)=0$, continues to scale linearly with system size $L$, as in the ordered VM, while the minority interaction markedly amplifies the velocity correlation amplitude (Fig.~\ref{fig:correlation}b and Fig.~\ref{fig:s4_correlation_length}). 

\begin{figure}
\centering
\includegraphics{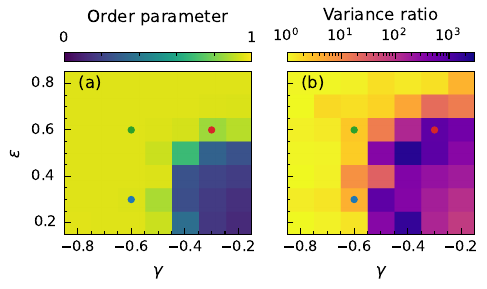}
\caption{\label{fig:heatmap} \textbf{Parameter dependence of avalanche behavior.} Heatmap showing (a) the time-averaged order parameter $\ev{\phi}$ and (b) the variance ratio $\text{Var}(\phi) / \text{Var}(\phi_{\text{VM}})$ across different values of minority interaction thresholds $\epsilon$ and $\gamma$. System parameters: $L=32, \rho=1.0, \eta=0.1$. A variance ratio of one corresponds to behavior identical to the standard VM, while higher values indicate stronger avalanche activity. We observe 10 to 1000 times higher variance over a large parameter range, indicating that the avalanche behavior does not require fine-tuning. For large $\gamma$ and small $\epsilon$ (lower right corner), avalanches become so frequent that they notably reduce the average order (a). The blue, green, and red dots correspond to the parameter combinations used for the lines of the same color in Fig.~\ref{fig:avalanche_stats}a,b.}
\end{figure}

To investigate the robustness of our findings across parameter space, we examine how the order parameter (Fig.~\ref{fig:heatmap}a) and the variance of the order parameter---which characterizes the magnitude and frequency of avalanche events---depends on the minority interaction parameters $\epsilon$ and $\gamma$ (Fig.~\ref{fig:s5_order_parameter} and Fig.~\ref{fig:s6_order_parameter_variance}). 
Fig.~\ref{fig:heatmap}b shows this variance normalized by the variance observed in the corresponding VM. 
Near the boundaries $\gamma=-1$ or $\epsilon=1$, the variance matches that of the standard VM because minority interactions are rarely triggered. 
However, in a substantial region of parameter space, we observe a significantly elevated variance---ten to 3000 times higher than in the VM---indicating that the avalanche dynamics are robust in the ordered phase rather than restricted to a narrow choice of $\epsilon$ and $\gamma$.
In the region of the parameter space where simultaneously $\gamma > -0.4$ and $\epsilon < 0.4$ the minority interaction is triggered so frequently that it prevents the formation of large scale order (Fig.~\ref{fig:heatmap}a and Fig.~\ref{fig:s5_order_parameter}).
We also find that global order decreases with system size when the minority interaction is present (Fig.~\ref{fig:s7_ordervssize}).
This is consistent with the formation and collision of several collectively migrating groups, which then trigger minority interactions.

\textit{Discussion}---A central question in collective animal behavior is:
How do individual behavioral rules combine to produce high responsiveness during collective motion? 
The macroscopic avalanches and strong velocity correlations across a broad parameter space observed here suggest that critical-like behavior and high responsiveness may emerge from the competition between majority alignment and minority interaction. 

This is consistent with recent theoretical work suggesting that heterogeneity and competing interactions can extend critical regions in complex systems \cite{sanchez-puig2023}.
Other theoretical works have also explored how deviations from pure consensus dynamics can drastically alter collective behavior. 
For instance, Bonilla and Trenado \cite{bonilla2019} studied a model where a fraction of agents act as systematic ``contrarians,'' orienting opposite to their neighbors. 
They found that this leads not to static order but to novel dynamical phases, including periodic oscillations and period doubling, demonstrating how minority opposition can prevent stable consensus and induce continual reorganization.
Codina et al. \cite{codina2022} investigated the effect of a fixed obstacle on the swarming dynamics in the VM.
They showed that an obstacle reflecting incoming particles leads to a globally disordered state of continuously interacting, locally ordered bands. 
These studies illustrate how complex alignment rules and geometry changes can prevent simple ordered states and drive complex macroscopic dynamics.  
However, none of these studies address our biological question of how biological swarms ensure responsiveness.

The scale-free velocity correlations and macroscopic avalanche dynamics in our model mirror observations across biological scales. 
These features closely parallel empirical observations in bird flocks \cite{cavagna2010, bialek2012a}, fish schools \cite{gomez-nava2023, handegard2012}, sheep \cite{ginelli2015}, and even collective cell migration \cite{chepizhko2016, wood2023, kawauchi2024, kim2024}.  
Notably, our minority interaction offers a mechanistic explanation for the remarkably rapid information transfer observed during collective predator evasion \cite{procaccini2011, gomez-nava2023, rosenthal2015}. 
When a small subset of individuals detects a threat, their deviating behavior can trigger an information cascade that propagates throughout the group at speeds far beyond the diffusion of direct individual interactions \cite{rosenthal2015}. 
These self-organized avalanches enable the emergent ``collective computation'' \cite{couzin2005} that gives animal groups their characteristic responsiveness to environmental perturbations while maintaining overall cohesion. 
Unlike traditional physical systems that require precise parameter tuning, biological collectives could take advantage of this competitive interaction mechanism to optimize information processing \cite{mora2011, klamser2021}.


Beyond biology, the minority interaction suggests a design principle for artificial swarms: responsiveness without central control or precise calibration \cite{khaluf2017}.
The balance between alignment and minority rules can be tuned for specific collective tasks.
The defector-based minority interaction analyzed here may also illuminate puzzling phenomena in public opinion. 
Classic social psychology shows that a consistent minority can reorient majority judgments \cite{moscovici1969}, while sociophysics models demonstrate that even a handful of deterministic contrarians destabilize otherwise robust majorities \cite{galam2004}. 
Psychologically, both a need for uniqueness \cite{lantian2017} and reactance to perceived constraints \cite{brehm1966, steindl2015} motivate individuals to endorse views precisely because they diverge from consensus. 
These findings suggest that in societies, like in our model, a small group of ``defectors'' can catalyze large-scale realignments in voting and belief landscapes.

Our results establish a minimal route to macroscopic reorientations and long-range correlations in collective motion in biologically relevant swarm sizes. 
The minority interaction endows the group with maximal responsiveness to localized cues while retaining cohesion, potentially offering a functional advantage in biological systems. 
Our work shows how biological collectives might balance reliability and flexibility using simple, local rules.
Due to its biological relevance we believe that this system warrants further investigation in the future, especially regarding the type of phase transitions observed, the universality class, asymptotic scaling behavior, and a possible analytical treatment. 

\begin{acknowledgments}
\textit{Acknowledgements}---SS and AD are funded by the European Union (ERC, subLethal, 101054921, https://erc.europa.eu/). 
Views and opinions expressed are however those of the authors only and do not necessarily reflect those of the European Union or the European Research Council Executive Agency. 
Neither the European Union nor the granting authority can be held responsible for them. 
CGR gratefully acknowledges funding from the Max Planck Society. 
JMNS acknowledges support from the PAPIIT-UNAM grant, project IN110726FP.
FP acknowledges financial support from INEX 2021 Ambition Project CollInt, projects ANR-22-CE30-0038 ``Push-pull" and (ANR-NSF) ANR-24-CE95-0002 ``MotDis".
The funders had no role in study design, data collection and analysis, decision to publish, or preparation of the manuscript. 
The authors thank the Department for Information Services and High Performance Computing (ZIH) at TUD Dresden University of Technology for providing excellent infrastructure.
\end{acknowledgments}
\bibliography{socswarming}

\appendix*\renewcommand{\thefigure}{S\arabic{figure}}\setcounter{figure}{0}
\section{Back Matter}
\textit{Model implementation}---As is standard for the VM, we use synchronous updates for the headings $\qty{\theta_i(t+\Delta t)}$, computed from the neighbor information at time $t$.
Both our VM and VM with minority interaction use Gaussian noise with zero mean and standard deviation $\sigma = \eta 2 \pi / \sqrt{12}$.
The model is implemented in Python 3.12, vectorized using numpy and optimized for speed using numba 0.61. 
We use a grid-based approach to nearest-neighbor search.

\begin{figure*}
\centering
\includegraphics{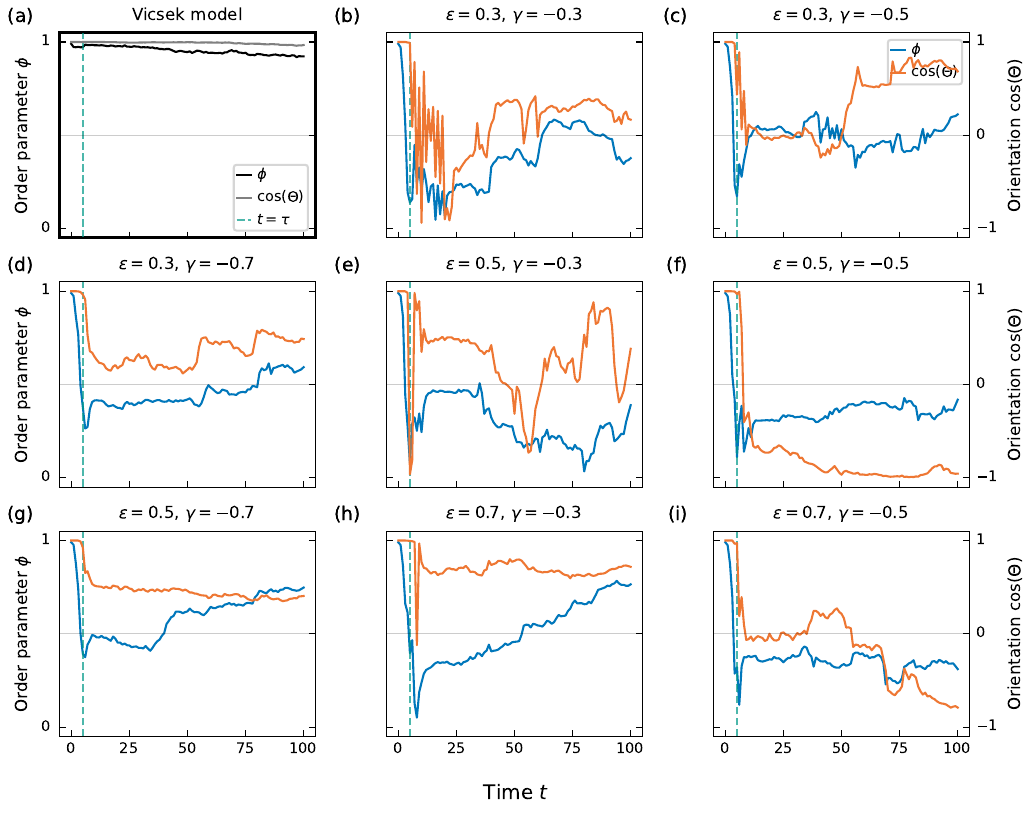}
\caption{\label{fig:s1_responsiveness} \textbf{Collective responsiveness to a controlled directional perturbation.}
Time evolution of the order parameter $\phi(t)$ (blue) and mean orientation $\cos(\Theta(t))$ (orange) after a single-particle perturbation: one particle is forced to orient at $\theta = \pi$ (opposite to the direction of a perfectly aligned flock; see Fig.~\ref{fig:minority_interaction}d-f for a sketch of the setting) for $\tau = 5$ time steps (dashed vertical line), then released.
(a) Standard Vicsek model.
(b)--(i) Model with minority interaction for eight combinations of $(\epsilon, \gamma)$.
The VM rapidly re-aligns with negligible directional change.
Across a broad range of minority interaction parameters, the localized perturbation is amplified into a group-wide reorientation event.
System parameters: $L=32$, $N=200$, $v_0=0.5$, $\eta=0.1$; single representative realizations.}
\end{figure*}

\textit{Responsiveness test}---To directly demonstrate enhanced collective responsiveness, we initialize an ordered flock and subject it to a brief, controlled directional perturbation: a single particle is forced to orient at $\theta = \pi$ (opposite to the collective direction) for $\tau = 5$ time steps, then released to resume normal dynamics (Fig.~\ref{fig:s1_responsiveness}).
The standard VM absorbs the perturbation with negligible change in order or global direction.
By contrast, across a broad range of $(\epsilon, \gamma)$ values the minority model can amplify this localized perturbation into a group-wide reorientation, consistent with the enhanced responsiveness reported in the main text.

\begin{figure*}
\centering
\includegraphics{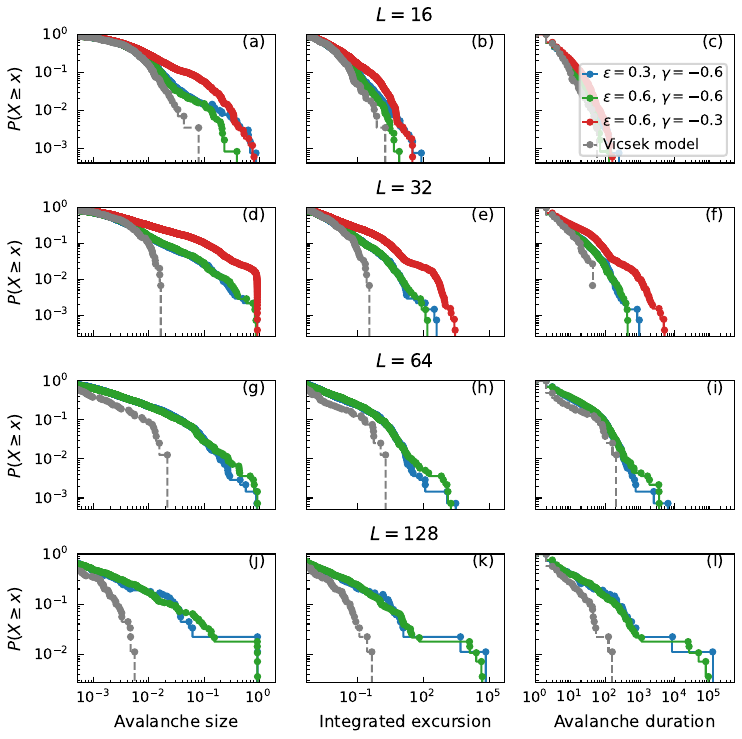}
\caption{\label{fig:s2_avalanches} \textbf{Avalanche statistics in dependence on system size.} Avalanche size (left column), integrated excursion (middle column) and avalanche duration (right column) for various system sizes (increasing from top to bottom) and threshold parameters (indicated by color). Integrated excursion and avalanche duration increase with system size spanning many orders of magnitude. Avalanches are always larger than in the corresponding VM (gray lines). Other parameters: $\rho=1, \eta=0.1$.}
\end{figure*}

\textit{Avalanche metrics}---Besides the avalanche size and avalanche duration defined in the main text we also define an integrated excursion
\begin{equation}
    S := \sum_{t \in \mathcal{I}} (\phi_c-\phi(t)) \Delta t,
\end{equation}
where $\mathcal{I}$ is the avalanche time interval, where $\phi \leq \phi_c$. 
The distribution of the integrated excursion $S$ varies over seven orders of magnitude and behaves otherwise similarly to avalanche size and duration -- indicating large-scale avalanches, see Fig.~\ref{fig:s2_avalanches} (middle column).

\textit{Code availability}---The code to reproduce the simulations shown in this paper is available as a Zenodo snapshot: \verb|10.5281/zenodo.18773616|.

\begin{figure*}
\centering
\includegraphics{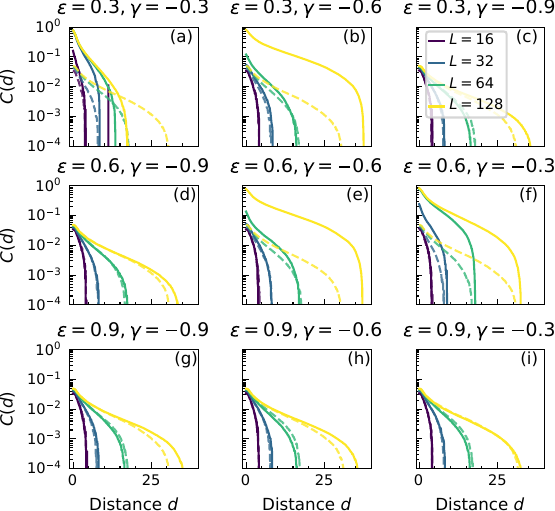}
\caption{\label{fig:s3_correlationfunction} \textbf{Velocity correlations in dependence on threshold parameters.} Spatial velocity correlation function (defined in Eq.~\ref{eq:correlation_func}) for nine combinations of threshold parameters (a-i) and system sizes (indicated by color) for the VM (dashed lines) and the VM with minority interaction (solid lines). When $\epsilon < 0.9 \lor \gamma > -0.9$ avalanches are triggered leading to strongly increased short-ranged velocity correlations (a, b, e, f). Other parameters: $L = 32, \rho=1, \eta=0.1$.}
\end{figure*}
\begin{figure*}
\centering
\includegraphics{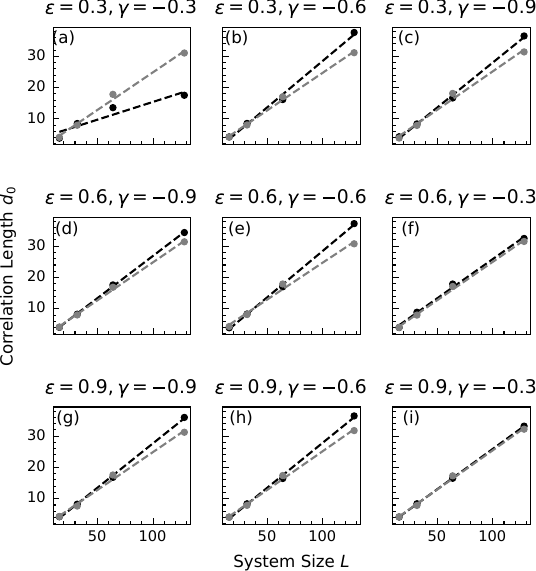}
\caption{\label{fig:s4_correlation_length} \textbf{Correlation length in dependence on threshold parameters.} Correlation length increases linearly for all combinations of threshold parameters. Black points indicate the model with minority interactions, gray points indicate the corresponding VM. If avalanches are triggered very easily (a) the correlation length is reduced. Otherwise, the correlation length in the VM with and without minority interaction is comparable in size. Dashed lines indicate linear fits. Other parameters: $L = 32, \rho=1, \eta=0.1$.}
\end{figure*}
\begin{figure*}
\centering
\includegraphics{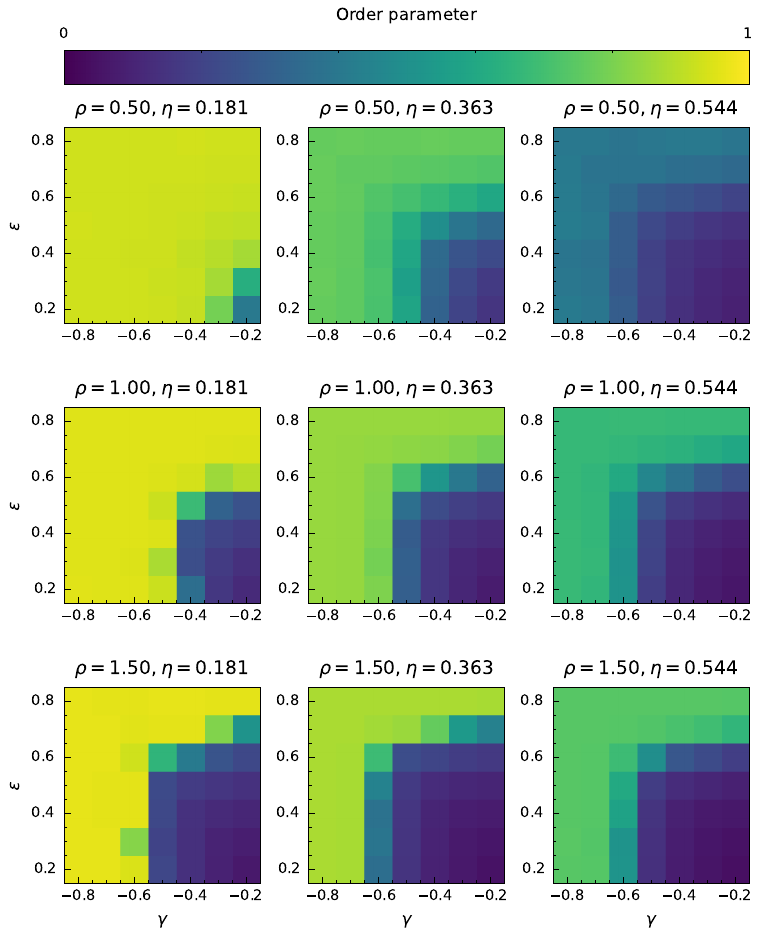}
\caption{\label{fig:s5_order_parameter} \textbf{Average order parameter in dependence on model parameters.} Heatmaps show the average order $\ev{\phi}$ in dependence on threshold values $\epsilon, \gamma$, density $\rho$ (increasing from top to bottom) and noise (increasing from left to right). Order increases with density and decreases with noise. For high $\gamma$ and low $\epsilon$ avalanches become very frequent and disrupt global order (lower right corner in the heatmap plots). System size: $L = 32$.}
\end{figure*}

\begin{figure*}
\centering
\includegraphics{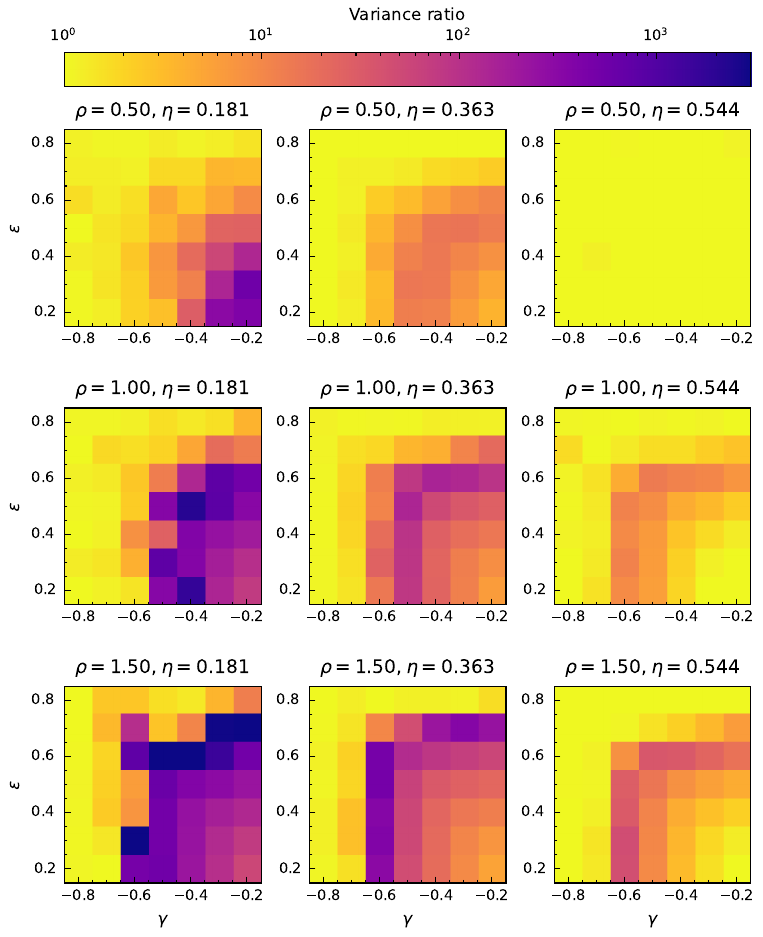}
\caption{\label{fig:s6_order_parameter_variance} \textbf{Variance of the order parameter relative to the VM in dependence on model parameters}. Heatmaps show the variance ratio $\text{Var}(\phi) / \text{Var}(\phi_{\text{VM}})$  in dependence on threshold values $\epsilon, \gamma$, density $\rho$ (increasing from top to bottom) and noise (increasing from left to right). In a large volume of the parameter space the order parameter fluctuates orders of magnitudes stronger than in the VM due to avalanches. For $\rho=0.5$ and $\eta=0.3$ (top right) the system is not in the ordered phase and behaves like the VM (compare also to the top right heatmap in Fig.~\ref{fig:s5_order_parameter}). System size: $L = 32$.}
\end{figure*}

\begin{figure*}
\centering
\includegraphics{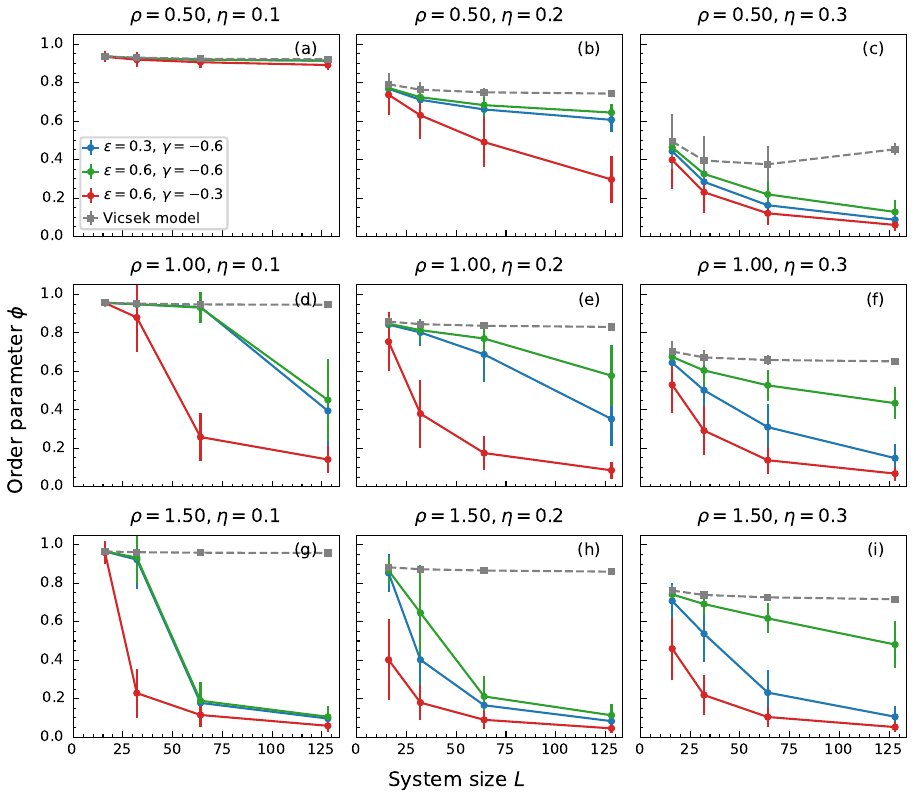}
\caption{\label{fig:s7_ordervssize} \textbf{Order decreases with system size}. Shown is the average order parameter $\ev{\phi}$ in dependence on system size $L$ for various combinations of threshold parameters (colored lines) and for the VM (gray dashed lines). Density increases from top (lowest) to bottom row (highest), while noise increases from left column to right column. In all cases the order parameter decreases as the system size is increased if the minority interaction is present, unlike in the VM.}
\end{figure*}

\end{document}